\documentstyle[12pt]{article}
\begin{document}

\begin{titlepage}
\begin{center}

\bigskip
\vspace{3\baselineskip}

{\Large \bf

$\Theta$-twisted Gravity \\}

\bigskip

\bigskip

{\bf 
Archil Kobakhidze \\}
\smallskip

{ \small \it Department of Physics and Astronomy, UNC-Chapel Hill, NC 27599-3255, USA \\ and \\
School of Physics, Research Centre for High Energy Physics \\ The University of Melbourne, Victoria 3010, Australia \\}

\bigskip

\vspace*{.5cm}

{\bf Abstract}\\
\end{center}
\noindent

We describe a theory of gravitation on canonical noncommutative 
spacetimes. The construction is based on $\theta$-twisted General Coordinate Transformations and Local Lorentz Invariance.

\bigskip

\bigskip

\end{titlepage}

\baselineskip=16pt

\paragraph{1.}
It can be suspected from various arguments, e.g.  related with short-distance divergences in quantum field theory and 
singularities in General Relativity, that the classical concept of spacetime continuum breaks down at small scales and must be replaced 
by some sort of 'quantum' spacetime with an intrinsic 'fundamental' lenght scale. In recent years, 
there were several attempts to formulate a theory of gravitation on ``quantized'' 
spacetimes \cite{Aschieri:2005yw}-\cite{Chamseddine:2000si}. The simplest such a spacetime is 
the spacetime with canonical noncommutativity\footnote{
We work with the Weyl-Moyal $\star$-product representation where noncommutative coordinates 
$\hat x^{\mu}$ (and their functions) are mapped to commutative coordinates $x^{\mu}$ with commutative 
pointwise product replaced by nonlocal $\star$-product:
$$f(\hat x)\to f(x);~f(x)\star g(x)=f(x)\exp\left(-\frac{i\ell^2}{2}\stackrel{\leftarrow}{P_{\mu}}\theta^{\mu\nu}
\stackrel{\rightarrow}{P_{\nu}}\right)g(x)~,$$
$P_{\mu}=-i\partial_{\mu}$ being the operator of spacetime translations.}
\begin{equation}
[x^{\mu}\stackrel{\star}{,}x^{\nu}]=i\ell^2\theta^{\mu\nu}~, 
\label{1}
\end{equation}  
where $\ell$ is a 'fundamental' length of noncommutativity and $\theta^{\mu\nu}=-\theta^{\nu\mu}$ are real constants. The $\star$-algebra of coordinates (\ref{1}) explicitly violates the basic symmetry of Einstein's General Relativity under the General Coordinate Transformations (GCT), leaving only 
its subgroup of (symplectic) volume-preserving diffeomorphisms. Noncommutative theory of gravity based on this residual symmetry 
has been constructed in \cite{Calmet:2005qm}, .
An important development in studies of (flat) noncommutative spacetimes was the identification 
of $\theta$-twisted Poncare group as a symmetry of noncommutative spacetimes \cite{Chaichian:2004za},\cite{Chaichian:2004yh}  (see also \cite{Oeckl:2000eg}). $\theta$-twisted diffeomorphisms has been introduced soon afterwards, and the corresponding theory of gravitation has bee
constructed in \cite{Aschieri:2005yw},\cite{Aschieri:2005zs}. In these works the Local Lorentz Invariance (LLI) 
is left unspecified since the authors essentially work with the second order formalism of General Relativity. 
However, LLI has to be consistently incorporated into a theory if one needs to couple fermions to gravity. 
Furthermore, the noncommutative metric tensor in \cite{Aschieri:2005yw},\cite{Aschieri:2005zs} is defined as 
a symmetrized $\star$-product of vierbeins, and this already requires specification of LLI, irrespective to the problem with fermions. 

In this paper we construct noncommutative theory of gravitation based on $\theta$-twisted symmetries of General Relativity, GCT and LLI, 
working in the first order formalism.  

\paragraph{2.}
An element of symmetry group of ordinary General Relativity can be represented as\footnote{This group has been introduced in 
\cite{Siegel:1978mj} as a starting point for their superspace formulation of supergravity.}:
\begin{equation}
G=\exp\left(i\xi^{\mu}(x)P_{\mu}+i\frac{1}{2}\lambda^{mn}(x)\Sigma_{mn}\right)~,
\label{2}
\end{equation}
where the first factor describes GCT\footnote{More precisely, these are GCT which can be continuously deformed to unity, known as a 'small gauge transformations' in gauge theory terminology. As far as one is not interested in global topological properties the small groups are sufficient to construct actions and derive corresponding local equations of motion.} and the second factor describes LLI. We use Greek indices for curved spacetime and those of Latin for flat vectors and tensors. The gauging of this group leads to the first order formulation of Einstein's General Relativity with vierbeins and spin connections being the gauge fields for GCT and LLI, respectively. A generic field $\Phi$ transforms according to a given representation $g$ as:
\begin{equation}
\delta \Phi=g\Phi.
\label{3}
\end{equation}
The transformation of the product of two fields then formally is given by a coproduct $\Delta(\delta)$:
\begin{equation}
\delta (\Phi_1\otimes \Phi_2)=\Delta(\delta)(\Phi_1\otimes \Phi_2)
\label{4}
\end{equation}
The coproduct has to be defined in the way to be compatible with the multiplication map $\mu$:
\begin{equation}
\mu\{\delta (\Phi_1\otimes \Phi_2)\}=\delta\mu\{\Phi_1\otimes \Phi_2\}~.
\label{5}
\end{equation}

For the ordinary commutative fields the coproduct $\Delta_0$ compatible with pointwise 
multiplication map $\mu_0\left(\Phi_1\otimes\Phi_2\right)=\Phi_1\Phi_2$ is given by:
\begin{equation}
\Delta_0(\delta)=\delta\otimes 1+1\otimes \delta
\label{6}
\end{equation}
This coproduct gives the usual Leibniz rule for the transformation of products of fields: $\delta(\Phi_1\otimes \Phi_2)=\delta\Phi_1\otimes \Phi_2+\Phi_1\otimes \delta\Phi_2$. 

To define a coproduct for noncommutative fields we first note that the $\star$-product map $\mu_{\star}$ can be 
expressed through the $\mu_0$ by twisting the tensor product of fields:
\begin{equation}
\Phi_1\star\Phi_2\stackrel{\rm def}{=}\mu_{\star}\{\Phi_1\otimes \Phi_2\}=\mu_0\{{\mathcal T}\Phi_1\otimes \Phi_2\}~,
\label{7}
\end{equation}
where ${\mathcal T}$ is the so-called twist operator defined as:
\begin{equation}
{\mathcal T}=\exp\left[-\frac{i\ell^2}{2}\theta^{\mu\nu}P_{\mu}\otimes P_{\nu}\right]~.
\label{8}
\end{equation}
Then it is easy to see that the compatibility condition (\ref{5}) is 
satisfied for the coproduct $\Delta_{\star}$ which is the ordinary coproduct $\Delta_0$ 
sandwiched by the twist operator ${\mathcal T}$ and its inverse ${\mathcal T}^{-1}=\exp\left[+\frac{i\ell^2}{2}
\theta^{\mu\nu}P_{\mu}\otimes P_{\nu}\right]$:
\begin{equation}
\Delta_{\star}={\mathcal T}^{-1} \Delta_0 {\mathcal T}~.
\label{9}
\end{equation}
Under the deformed coproduct (\ref{9}) the Leibniz rule is deformed as well:
\begin{eqnarray}
\delta (\Phi_1\star \Phi_2)=\mu_0\{ \Delta_0 {\mathcal T}(\Phi_1\otimes \Phi_2) \}=
\delta(\Phi_1)\Phi_2+\Phi_1\delta(\Phi_2)- \\
\sum_{n=1}^{\infty}\frac{(i\ell^2)^n}{2^nn!}\theta^{\alpha_1\beta_1}...
\theta^{\alpha_n\beta_n}\left[(\delta P_{\alpha_1}...P_{\alpha_n}\Phi_1)(P_{\beta_1}...P_{\beta_n}\Phi_2)+
(P_{\alpha_1}...P_{\alpha_n}\Phi_1)(\delta P_{\beta_1}...P_{\beta_n}\Phi_2)\right]~.\nonumber
\label{10}
\end{eqnarray}
GCT and LLI defined through the above deformation of coproduct we call $\theta$-wisted symmetries upon which the noncommutative 
gravity should  be constructed. In particular, note that the $\star$-commutator of coordinates (\ref{1}) is 
invariant under the $\theta$-twisted GCT:
\begin{equation}
\delta_{\xi}[x^{\mu}\stackrel{\star}{,}x^{\nu}]=\delta_{\xi}[x^{\mu},x^{\nu}]=0~.
\end{equation} 

\paragraph{3.}
Now we proceed with gauging the $\theta$-twisted symmetries defined above. Following the standard route we introduce the gauge potential
\begin{equation}
A_m=ie^{\mu}_mP_{\mu}+i\frac{1}{2}\omega_m^{~kl}\Sigma_{kl}~,
\label{11}
\end{equation} 
where $e^{\mu}_m(x)$ is a (inverse) vierbein - the gauge field for GCT, and 
$\omega_m^{~kl}(x)$ is a spin connection - the gauge field for LLI. A peculiar 
point for General Relativity is that the above gauge potential plays a dual role. 
Namely, it represents the covariant derivative at the same time. 
Thus the field strengths can be computed straightforwardly:
\begin{equation}
i[A_m\stackrel{\star}{,}A_n]=\frac{1}{2}R_{mn}^{~~kl}\Sigma_{kl}+T_{mn}^{~~\mu}P_{\mu}~,
\label{12}
\end{equation}
where
\begin{equation}
R_{mn}^{~~kl}\Sigma_{kl}=(A_m\star \omega_n^{~~kl}-A_n\star \omega_n^{~~kl})\Sigma_{kl}
\label{13}
\end{equation}
is the field strength for LLI (the analogue of the Riemann tensor), and 
\begin{equation}
T_{mn}^{~\mu}=A_m\star e^{\mu}_n - A_n\star e^{\nu}_m~. 
\label{14}
\end{equation}
is the field strenght for GCT (torsion).

Let us show now that all these objects indeed transform covariantly under the $\theta$-twisted symmetries. 
The gauge potential (\ref{11}) transforms  
as an adjoint representation as usually \cite{Vassilevich:2006tc}, \cite{Aschieri:2006ye}: 
$A_m^{\prime}(x)={\rm e}^{i\Lambda}A_m{\rm e}^{-i\Lambda (x)}$, where $\Lambda(x)=
\xi^{\mu}(x)P_{\mu}+\frac{1}{2}\lambda^{kl}(x)\Sigma_{kl}$. 
 For the infinitesimal transformations we obtain:
\begin{equation}
\delta A_m=i[\Lambda, A_m]~.
\label{15}
\end{equation}
The eqn. (\ref{15}) has exactly the same form as in the commutative case. 
Then using the proper coproduct (\ref{9}) it is straightforward to see that (\ref{12}) is covariant:
\begin{equation}
\delta [A_m\stackrel{\star}{,}A_n ]=i[\Lambda,[A_m\stackrel{\star}{,}A_n ]]
\label{16}
\end{equation} 
Since the torsion is covariant under the $\theta$-twisted symmetries, we can set it to zero,
\begin{equation}
A_m\star e^{\mu}_n =A_n\star e^{\nu}_m
\label{17}
\end{equation}
This gives the relation between the spin connection and vierbein. Of course this relation is deformed relative to the 
commutative torsion-free condition because of the twist. From the covariant Riemann tensor we can construct noncommutative 
Ricci tensor and Ricci scalar in the usual way:
\begin{equation}
R_{m}^{~n}=R_{mk}^{~~kn},~~R=R_{m}^{~m}~.
\label{18}
\end{equation}
The action for the noncommutative General Relativity then reads:
\begin{equation}
S_{\rm NCGR}=\int d^4x {\bf \rm e}^{\stackrel{\star}{-}1}R~,
\label{19}
\end{equation}
where we have removed the $\star$-product by droping full derivative term. 
In eqn (\ref{19}) ${\bf \rm e}^{\stackrel{\star}{-}1}$ is the $\star$-inverse of the $\star$-determinant  ${\bf \rm e}$,
\begin{equation}
 {\bf \rm e}=\frac{1}{4!}\varepsilon^{mnkl} 
\varepsilon_{\mu\nu\rho\sigma}e_m^{\mu}\star e_n^{\nu}\star e_k^{\rho}\star e_l^{\sigma}~,
\label{a}
\end{equation}
\begin{equation}
 {\bf \rm e}\star{\bf \rm e}^{\stackrel{\star}{-}1}=1
\label{b}
\end{equation}
The $\star$-inverse can be expressed through the ordinary commutative inverse ${\bf \rm e}^{-1}$ by means of geometric series 
\cite{Aschieri:2005yw}:
\begin{equation}
{\bf \rm e}^{\stackrel{\star}{-}1}= {\bf \rm e}^{-1}\star\sum_{k=0}^{\infty}(1- {\bf \rm e}\star {\bf \rm e}^{-1})^{\star k}=
{\bf \rm e}^{-1}+{\bf \rm e}^{-1}\star(1-{\bf \rm e}\star {\bf \rm e}^{-1})+...
\label{c}
\end{equation}
Using the twisted coproduct of eqn. (\ref{9}) one can easily verify that the $\star$-inverse transforms in 
the same way as the commutative inverse. Hence the action 
(\ref{19}) is indeed invariant under the $\theta$-twisted GCT and LLI. More explicit treatment of the proposed theory with some physical applications will be given in forthcoming publications. 
 
\paragraph{4.}
Concluding, we have constructed a theory of gravitation on cannonical noncommutative spacetimes. The key ingredients of our construction is the Lie-algebraic approach to General Relativity, where GCT and LLI are treated as gauge symmetries. While the classical GCT is broken in noncommutative spacetime, we demonstrate that (\ref{1}) is invariant under the $\theta$-twisted GCT. Thus the 'true' symmetries of noncommutative General Relativity are $\theta$-twisted GCT and LLI. The action (\ref{19}) obtained by gauging $\theta$-twisted GCT and LLI is invariant under these symmetries. 

Perhaps the most interesting aspect of spacetime noncommutativity is the possible relevance of spacetime fuzziness to the problem of 
singularities. It would be interesting to study the deformations of classical Schwarzschild and cosmological solutions within the 
noncommutative General Relativity proposed in this work.

\subparagraph{Acknowledgments.}
This work was completed while I was visiting the CCPP, New York University. I would like to thank Gia Dvali and Gregory Gabadadze for discussions.

\end{document}